# Unipolar Time-Differential Pulse Response with a Solid-State Charpak Photoconductor

A. H. Goldan[a,1], O. Tousignant[b], K. S. Karim[c], J. A. Rowlands[d]

[a] *Department of Radiology, School of Medicine, SUNY at Stony Brook, NY, US*
[b] *ANRAD Corporation, 4950 Levy Street, Saint-Laurent, PQ, Canada*
[c] *Electrical and Computer Engineering, University of Waterloo, Waterloo, ON, Canada*
[d] *Thunder Bay Regional Research Institute, Thunder Bay, ON, Canada*

## ABSTRACT

**We demonstrate a high granularity multi-well solid-state detector with the unipolar time-differential property. Results show an improvement in the temporal pulse response by more than two orders-of-magnitude using amorphous selenium as the photoconductive film. The significance of the results presented here is the ability to reach the intrinsic physical limit for detector pulse speed by transitioning from the slow transit-time-limited response which depends on the bulk carrier transport mechanism, to the ultrafast dispersion-limited response which depends on the spatial spreading of the collected carrier packet.**

[1] E-mail: Amirhossein.Goldan@stonybrookmedicine.edu

Soon after the Nobel prize winning invention of the gas-filled multiwire proportional chamber (MWPC) by Charpak[1], and parallel to developments in microelectronics, a great deal of research was stimulated to develop the highest *granularity*[2] gaseous detectors for achieving the highest position resolution. However, the practical benefits of high granularity was restricted by the micro- to milli-metre range of the radiation induced photoelectron cloud in gas. Solid materials, on the other hand, can have three orders-of-magnitude shorter photoelectron range due to much higher density, and thus, they yield much smaller detector dimensions with substantially higher spatial and temporal resolution[3]. The problem is that disordered solids, which are easier and less expensive to develop than single crystalline solids, have been ruled out as viable radiation detection media because of slow pulse response, which is attributed to their low carrier mobilities and transit-time-limited photoresponse.

In this letter, we show a high granularity multi-well solid-state detector (MWSD) fabricated using photolithography and film evaporation techniques[4]. We directly probe the transit of photoinduced carriers in the MWSD using time-of-flight (TOF) transient photoconductivity measurements and show time-differential responses to optical laser and x-ray excitations.

Let's consider a medium within which a single drifting excess carrier is contained. According to the Ramo-Shockley theorem[5], the induced current on the collector due to carrier displacement inside the medium is given as

$$I_i(t) = q\frac{\partial V_W}{\partial t} = q\mu F\frac{\partial V_W}{\partial z} \tag{1}$$

where $q$ is the charge of the moving carrier, $\mu$ is the effective carrier mobility inside the medium, $F$ is the applied external field, $V_W$ is the *weighting potential*[6], and $z$ is the carrier displacement during time $t$ which is $z = \mu F t$. Note that the conceptual $V_W$ is dimensionless and is the potential that would exist in the detector with the corresponding collecting electrode raised to unity and all other electrodes grounded.

Conventionally, a planar photoconductive material is fitted between two parallel contacts to form a sandwich cell: one is the drift electrode that is kept at a certain potential, $V_D$, with proper polarity and magnitude, and the second is the collecting electrode (or the collector) that is generally biased at zero potential and connects to the readout electronics for signal capture. The weighting potential for these parallel plate detectors (PPDs) is zero at the drift electrode and rises linearly to one at the collector (dashed line in Fig. 1). According to Eq. 1, such distribution means that the collector is sensitive to real-time bulk transport, and thus, pulse response is limited by the carrier transit-time, $t_T$, across the photoconductor thickness, $L$ (top inset plot in Fig. 1)

$$I_{i1}(t) = \frac{q}{t_T}(H(t) - H(t - t_T)) \tag{2}$$

where $H(t)$ is the Heaviside step function and $t_T = L/\mu F$ for an excess carrier drifting across the detector thickness. Equation 2, with its well-defined plateau, assumes: (a) medium homogeneity, (b) coherent carrier drift with non-dispersive Gaussian transport properties, (c) homogeneous field distribution, (d) $RC \ll t_T \ll \tau_D$, and (e) $t_T \ll \tau_{rel}$, where $RC$ is the time constant of the readout circuit, $\tau_D$ is the deep-trapping lifetime, and $\tau_{rel}$ is the dielectric relaxation time.

Now consider a new device with its weighting potential at zero everywhere in the bulk except for a very small region near the collector where it rises sharply to one (solid line in Fig. 1). In this case, the induced photocurrent due to a single carrier drift is an impulse (bottom inset plot in Fig. 1)

$$I_{i2}(t) = q\delta(t - t_T) \propto \frac{\mathrm{d}}{\mathrm{d}t} I_{i1}(t)\bigg|_{t=t_T} \tag{3}$$

where $\delta(t)$ is the impulse function. In this case, the collector is completely insensitive to bulk event times of the drifting excess carrier and photoresponse is independent of photoconductor material (whether single crystalline or disordered, organic or inorganic) and its carrier transport mechanism (whether coherent or incoherent drift via band transport, multiple-trapping, or hopping). Note that the shape of $I_{i2}$ is the time-derivative of $I_{i1}$ at $t_T$ and that rise time at the collector is only limited by the $RC$ time constant of the readout electronics.

To circumvent the problems that originate from poor bulk transport properties in disordered solids, one must use the proposed device with its modified $V_W$ distribution by decoupling the radiation absorbing photoconductor from the collector. In bipolar solids such as chalcogenide glass amorphous selenium (a-Se), where charge induction is due to the drift of both electrons ($-$) and holes ($+$) in a conventional PPD, this decoupling has the added advantage of sensing only the carrier type with a higher mobility-lifetime product (i.e., the primary carrier) and providing insensitivity of the collector to the transport properties of the slower carrier type. Such localized preferential sensing of primary carriers, which is also referred to as unipolar (or single-polarity) charge sensing, can be implemented by (1) proper potential biasing of the sandwich electrodes to drift primary carriers towards the collector, and (2) establishing a strong *near-field effect* in the immediate vicinity of the collector using an *electrostatic shield*. The near-field effect can be established with either a direct approach using the Frisch grid design[7,8] or the small-pixel effect[6], or an indirect approach using the coplanar pixel electrodes[9].

Inspired by Charpak's MWPC[1] and its micropattern variants[7,8], we have fabricated a solid-state detector with an internal electrostatic shield using grid-on-insulator (GOI) and a-Se film evaporation techniques[4]. The proposed device is called the multi-well solid-state detector (MWSD) and its structure is shown with schematic in Fig. 2a and with scanning electron microscope (SEM) cross-section in Fig. 2b. The device consists of an evenly spaced insulating pillars over the collector with pillars' top side coated with a conductive layer to form the shielding grid. Note that for simplicity of fabrication, the insulator over the collector is not etched, which inhibits neutralization of the drifting charge by the collector. However, we have limited the build-up of this surface space charge using low-level excitation for a rested specimen, and thus, space charge perturbation is minimized.

The weighting potential distribution of the fabricated device is simulated in Fig. 2c which shows a very small change in $V_W$ in the region between the drift electrode and the grid (i.e., the interaction region). However, $V_W$ changes substantially inside each well in the region between the grid and the collector (i.e., the detection region). Similar to the Frisch gas chambers[7,8] and the coplanar detectors[9], we must bend the electric field lines in the drift volume close to the grid so that all primary carriers are steered away from the grid and channeled inside the well towards the collector. Thus, we define the field bending ratio $r = V_G/V_D$, where $V_G$ is the grid bias. For comparison purposes, a conventional PPD without the shielding grid was also fabricated on the same substrate and Fig. 2c shows its linear $V_W$ distribution.

We used the time-of-flight (TOF) transient photoconductivity measurements[10,11,12,13] to verify our theoretical time-differential prediction in Eq. 3. For all the TOF experiments reported in this paper, we operated the devices at room temperature with a bulk field of $F = 2$ V/$\mu$m and $r = 0.2$ (i.e., for PPD: $V_D = 400$ V, for MWSD: $V_D = 500$ and $V_G = 100$ V). Also, we considered more realistic experimental impulse-like excitations, in contrast to a single drifting excess carrier, where (1) a sheet of carriers is photoinduced close to the drift electrode with a blue laser pulse and (2) Gaussian carrier clouds are generated uniformly across the bulk with a high energy x-ray pulse. Emphasis is on measured results obtained from chalcogenide glass a-Se but conclusions drawn can be extended to other non-dispersive inorganic and organic photoconductive materials because of the universal feature of charge transport that is independent of their atomic, molecular, and crystalline structures. For example, in the case of non-dispersive Markoffian transport, Scher-Montroll (SM) *universality* of the photocurrent is not applicable and the propagating carrier packet experiences spreading which is described by Gaussian statistics[14], where the position of the peak of carrier distribution coincides with its spatial mean (Fig. 3a). This spreading is mainly due to fluctuations of the shallow-trap release time and, in the small-signal case[15], we can neglect spreading due to mutual Coulomb repulsion of the free charge density. The Gaussian statistics for the root-mean-square (rms) drift spread, $\sigma_D$, and the mean carrier displacement, $\ell$, obey time dependencies $\sigma_D \propto t^{1/2}$ and $\ell \propto t$, which yield the well-known relation $\sigma_D/\ell \propto t^{-1/2}$ (Ref.[14]).

For optical TOF experiments, we used a VSL337 dye laser tuned to 337 nm wavelength (for strong absorption in a-Se) and 5 ns pulse duration (for impulse-like excitation). Figure 3b shows non-dispersive hole photocurrent transients in a-Se. The result obtained from the PPD shows a semi-rectangular pulse with a soft plateau, due to inhomogeneous field distribution, followed by an exponential decay which is the Gaussian integral of the total drift spread. However, the MWSD response shows a Gaussian pulse (centered at the hole transit time $t_{T+}$) that verifies the time-differential property of Eq. 3. The time-differential Gaussian TOF, which is similar to a typical time distribution of Charpak's MWPC (Fig. 29 in Ref.[16]), signifies the ability to reach the intrinsic physical limit for pulse speed in non-dispersive solids.

For high energy penetrating radiation, photon interaction can occur throughout the bulk which results in depth-dependent signal waveform variations in PPDs[6]. However, signal waveforms are depth-independent and unipolar in the MWSD because only holes drift through the multi-well and are sensed by the collector. Thus, the detector pulse speed is improved substantially by a factor of $n \leq (L/4\sigma)(\mu_+/\mu_-)$, where the first term is due to the time-differential property, the second term is due to the unipolarity, $\sigma$ is the total rms spread, and $4\sigma$ is approximately the Gaussian pulse width. As shown in Fig. 3c, $n$ is equal to $\sim$300 for a-Se comparing hole-dispersion-limited response with that of electron-transit-time-limited.

The next experiment extends the concept of optical TOF to x-rays that is also applicable to a blocking drift electrode[13]. The x-ray TOF is different from optical TOF in that (1) photon absorption can occur throughout the photoconductor, and (2) a Gaussian charge-cloud is formed around the primary interaction site of each absorbed photon (inset of Fig. 4). The rms spreading $\sigma_R$ of the charge cloud (i.e., the photoelectron range) obeys the relation $\sigma_R \propto E^2$ in a-Se[17], where $E$ is the energy of the absorbed x-ray photon. For x-ray excitations, we used an XR200 pulsed source tuned to 150 kvP (for a nearly uniform charge-cloud generation density across the photoconductor thickness), 3 mR exposure (for maintaining the small-signal case), and 60 ns pulse duration (for impulse-like excitation). The measured time-resolved transients in Fig. 4 show a linear decay (i.e., triangular response) for the conventional PPD due to carrier neutralization at the collector (or in our case, carrier immobilization at the Se-PI interface), and a nearly constant response (i.e., rectangular) for the proposed MWSD, verifying once again the time-differential property. An important feature of the x-ray time-differential response is the observed exponential tail, representing the actual Gaussian distribution of the last drifting hole packet that was initially generated close to the drift electrode from an absorbed x-ray photon with $E = 60$ keV. This response at the tail shows the physical limit for the photoconductor's temporal performance due to the total spatial spreading $\sigma = \sqrt{{\sigma_D}^2 + {\sigma_R}^2} = 0.07$. Note that the spike observed at the onset of the x-ray pulse is the result of surface space-charge perturbation. The magnitude of this spike is larger for the MWSD because of its higher field in the detection region (Fig. 4 with the top-right axes).

An important non-ohmic effect in disordered solids may occur in the presence of a strong field with the transport mechanism shifted from localized states into extended states where the mobility can be 100 to 1000 times higher[18]. Such hot carriers in extended states (with mobilities near the mobility edge) can gain energy faster than they lose it to phonons, and thus, avalanche due to impact ionization is possible (e.g., hot holes in a-Se[19] in contrast to hot electrons in amorphous silicon[20]). Continuous and stable avalanche multiplication has been shown in a-Se, a feature that enabled the development of an optical camera with more sensitivity than the human eye (i.e., 11 lx at aperture F8, or 100 times more sensitive than a CCD camera)[21]. For high-energy penetrating radiation, the challenge is that avalanche-mode selenium cannot be the bulk medium because (1) avalanche layers cannot be very thick ($< 25\mu$m) and (2) a uniform avalanche field in the bulk causes depth-dependent gain variations. Our proposed MWSD (Fig. 2b) is the practical approach for achieving stable avalanche in large-area direct radiation detectors, where the low-field interaction region can be made as thick as necessary to stop high-energy radiation, and the high-field multi-well detection region can be optimized for avalanche multiplication.

In conclusion, we have designed, fabricated, and characterized a high granularity multi-well solid-state detector which achieves ultrafast unipolar time-differential pulse response. It is important to remark that we were able to reach the physical limit of pulse speed in a-Se set by the spatial spreading of the collected hole cloud. Future studies are envisaged to build devices with proper blocking contacts for achieving avalanche-multiplication gain in the wells, with applications ranging from high-energy and nuclear physics to industrial and medical diagnostics and crystallography[22,23]. Furthermore, advances in nano-electronics can be applied to manufacture the highest granularity nanopattern solid-state detectors (NSSD) with ultrafast time-differential photoresponse, due to the nanoscale photoelectron range in solids in response to impulse excitations in a continuum from the visible to the soft x-rays, with applications in optical communications[24] and time-domain spectroscopy[25].

## Acknowledgements

We are grateful to Dr. Giovanni DeCrescenzo for help with the characterization of the detectors. Multi-well structures were fabricated photolithographically using the Giga-to-Nanoelectronics Centre facilities at the University of Waterloo and amorphous selenium films were deposited at ANRAD corporation.

## Figure Captions

**Figure 1**: Weighting potential distributions for a conventional PPD (dashed line) and a new hypothetical detector (solid line). Insets show the corresponding induced photocurrents due to a single excess carrier drift.

**Figure 2**: **(A)** Schematic diagram for our realization of the proposed device, called the multi-well solid-state detector (MWSD). **(B)** SEM cross-sections of the representative device. The polyimide (PI) pillars are 11.5 $\mu$m in height and are evenly distributed over the chromium (Cr) collector with a pitch of 10 $\mu$m. Their top surface is covered with Cr to form the shielding grid electrode. To ensure blocking contacts for limiting the excess charge injection, the grid is coated with another thin PI layer. A 200 $\mu$m a-Se film is evaporated over the structure as the photoconductive material (i.e., $L = 200$ $\mu$m), and finally, a semi-transparent gold (Au) layer is sputtered on top to provide the drift electrode while enabling optical excitation measurements. The top contact is non-blocking not to impede the extraction of optically induced carriers.**(C)** Weighting potential distributions for carriers terminating on the collector and on the grid.

**Figure 3**: **(A)** Schematic representation of carrier packet transport optically induced close to the drift electrode in a non-dispersive solid. **(B)** Hole TOF transients in a-Se PPD and MWSD structures. Gaussian TOF transient in the MWSD with $(\sigma_D/\ell)_{t_{T+}}$=0.04, shows the time-differential property. **(C)** Logarithmic TOF plots showing the signal pulse widths. For an x-ray photon absorbed close to the collector, holes are immediately neutralized and only electrons are in motion towards the drift electrode. Thus, pulse speed is limited by the electron transit time $t_{T-}$ in a-Se PPDs, as shown from the measured electron TOF. Measured a-Se effective electron and hole mobilities are $\mu_- = 0.002$ and $\mu_+ = 0.1$ cm$^2$ V$^{-1}$ s$^{-1}$, respectively.

**Figure 4**: Bottom-left axis: Linear-decay x-ray TOF response of the PPD and unipolar rectangular response of the MWSD, showing once again its time-differential pulse response. The measured total spreading of the Gaussian tail is $(\sigma/\ell)_{t_{T+}}$=0.07. Top-right axis: Electric field distribution in the bulk. Top inset shows schematic representation of Gaussian photoelectron clouds created at the onset of radiation ionization.

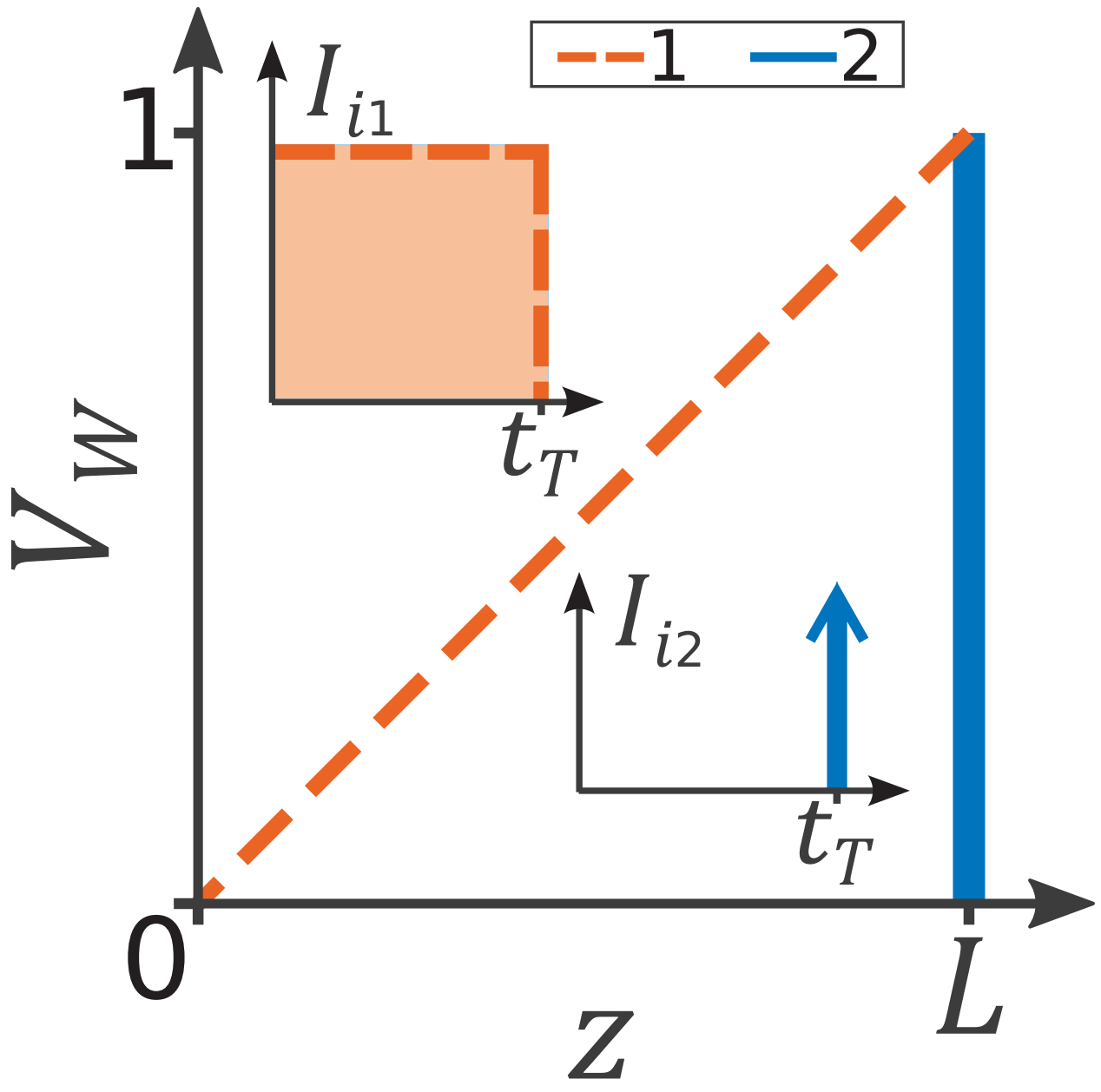

FIGURE 1

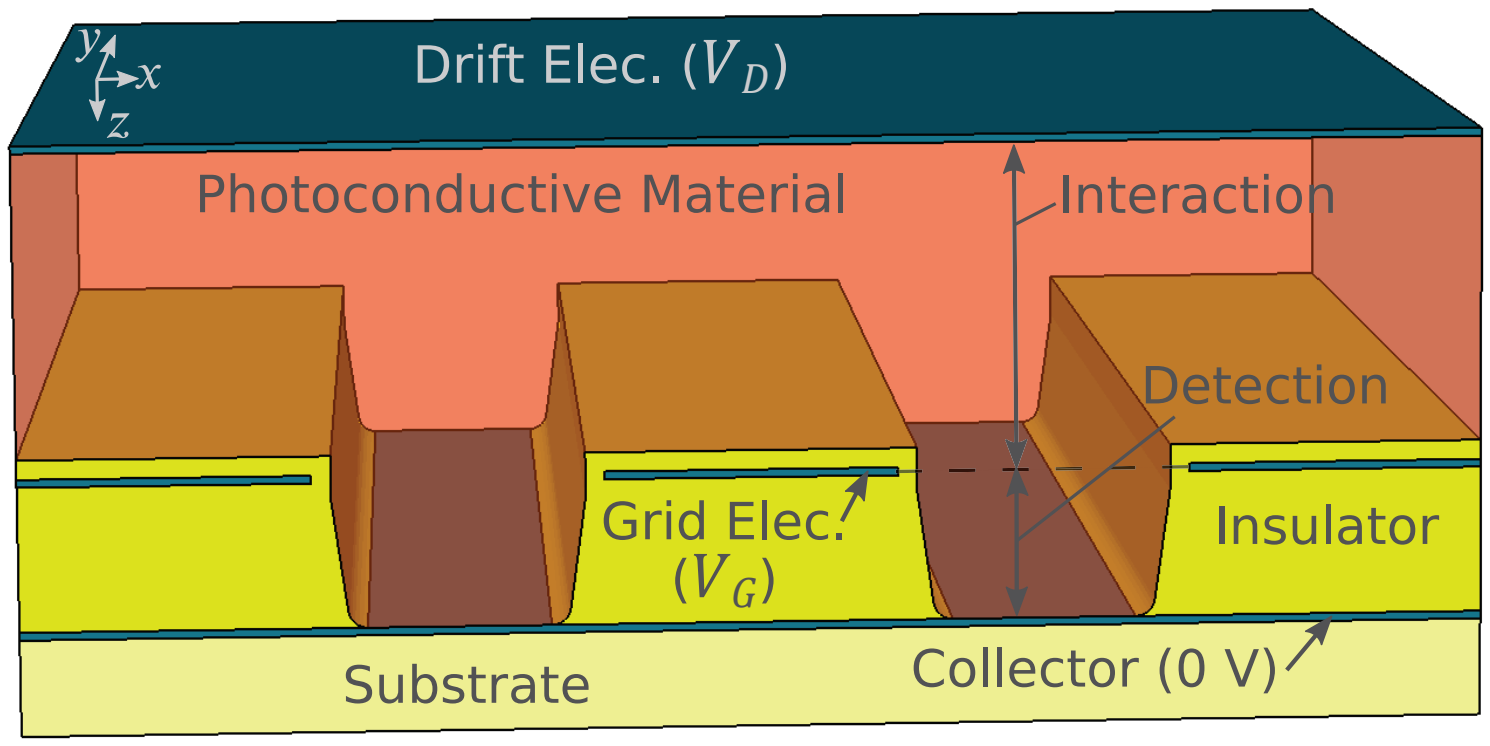

(A)

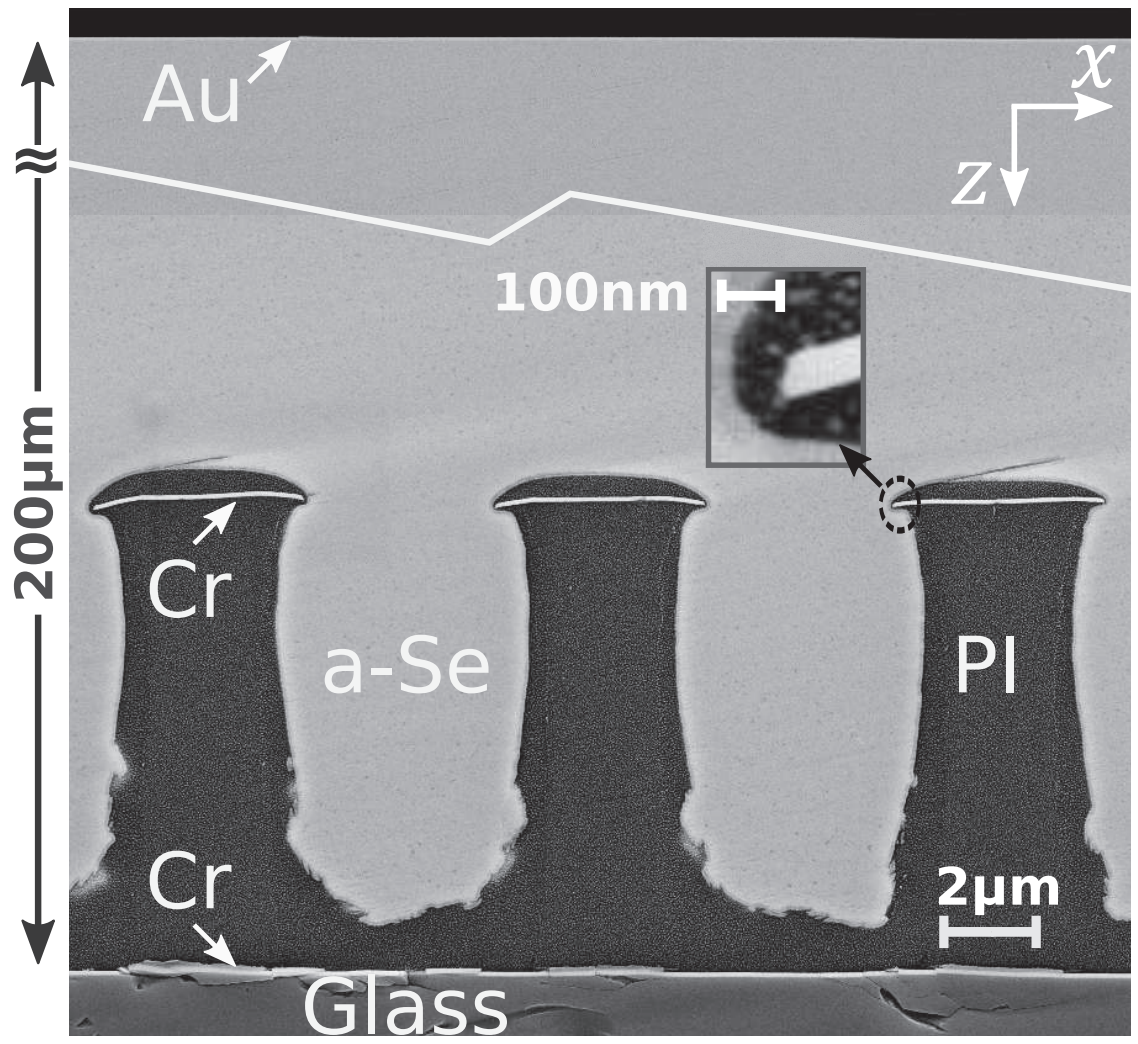

(B)

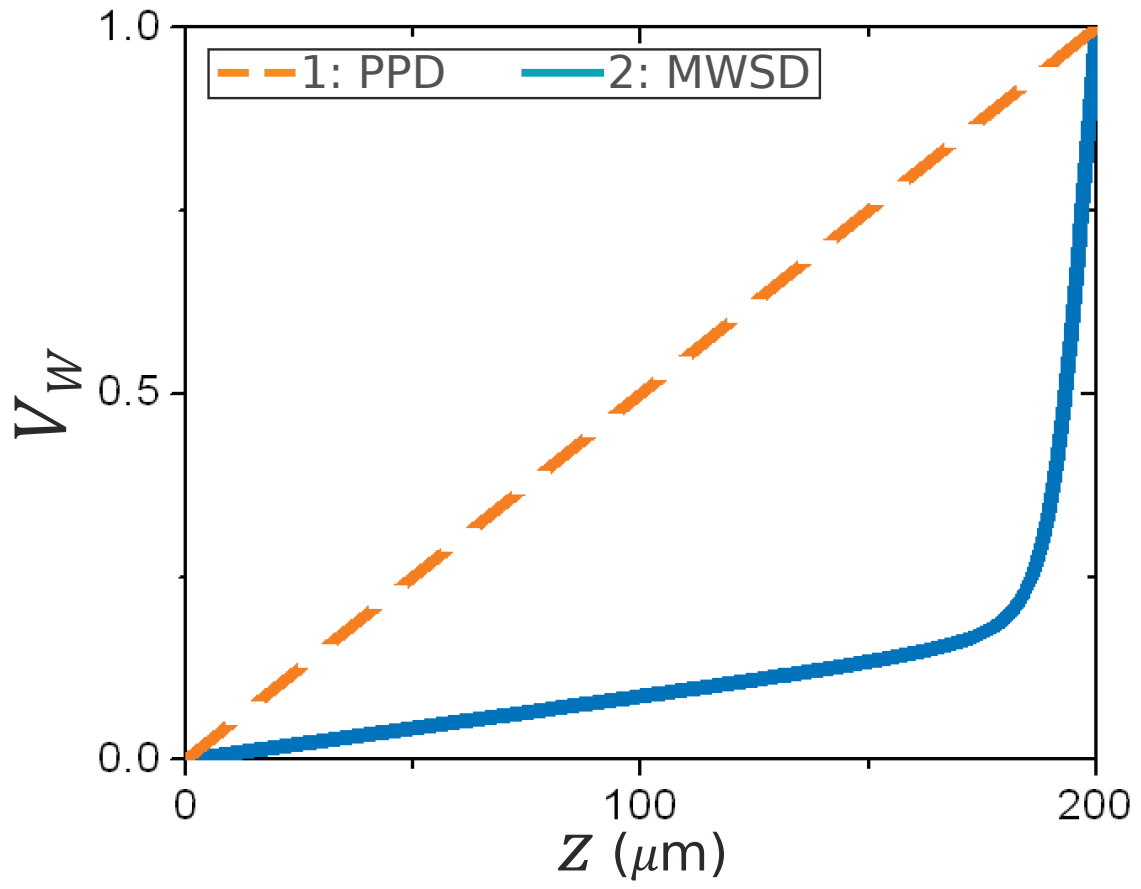

(C)

FIGURE 2

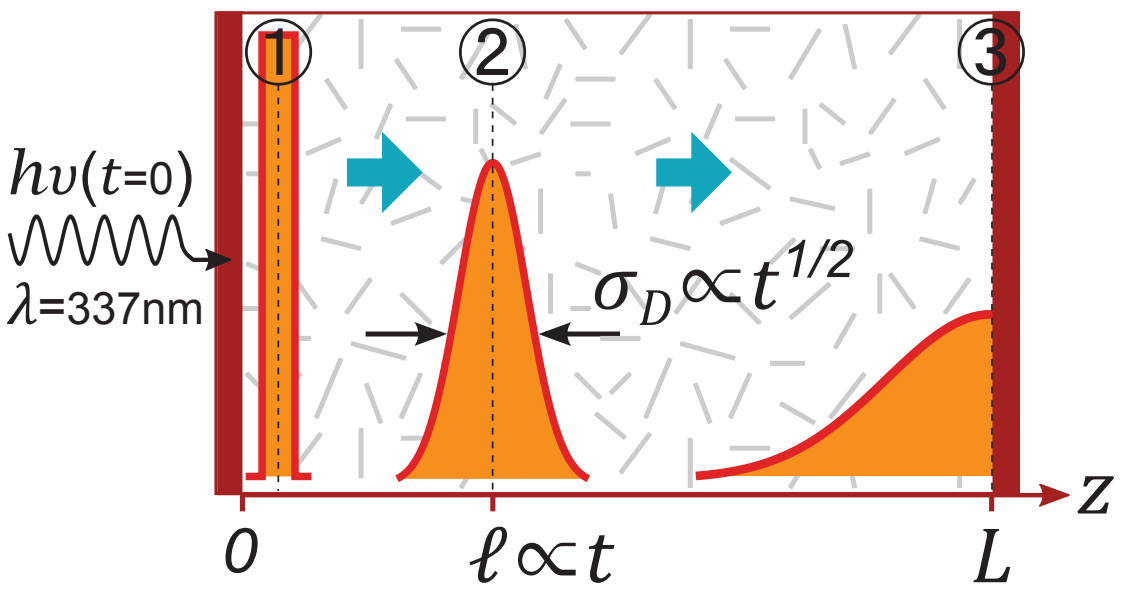

(A)

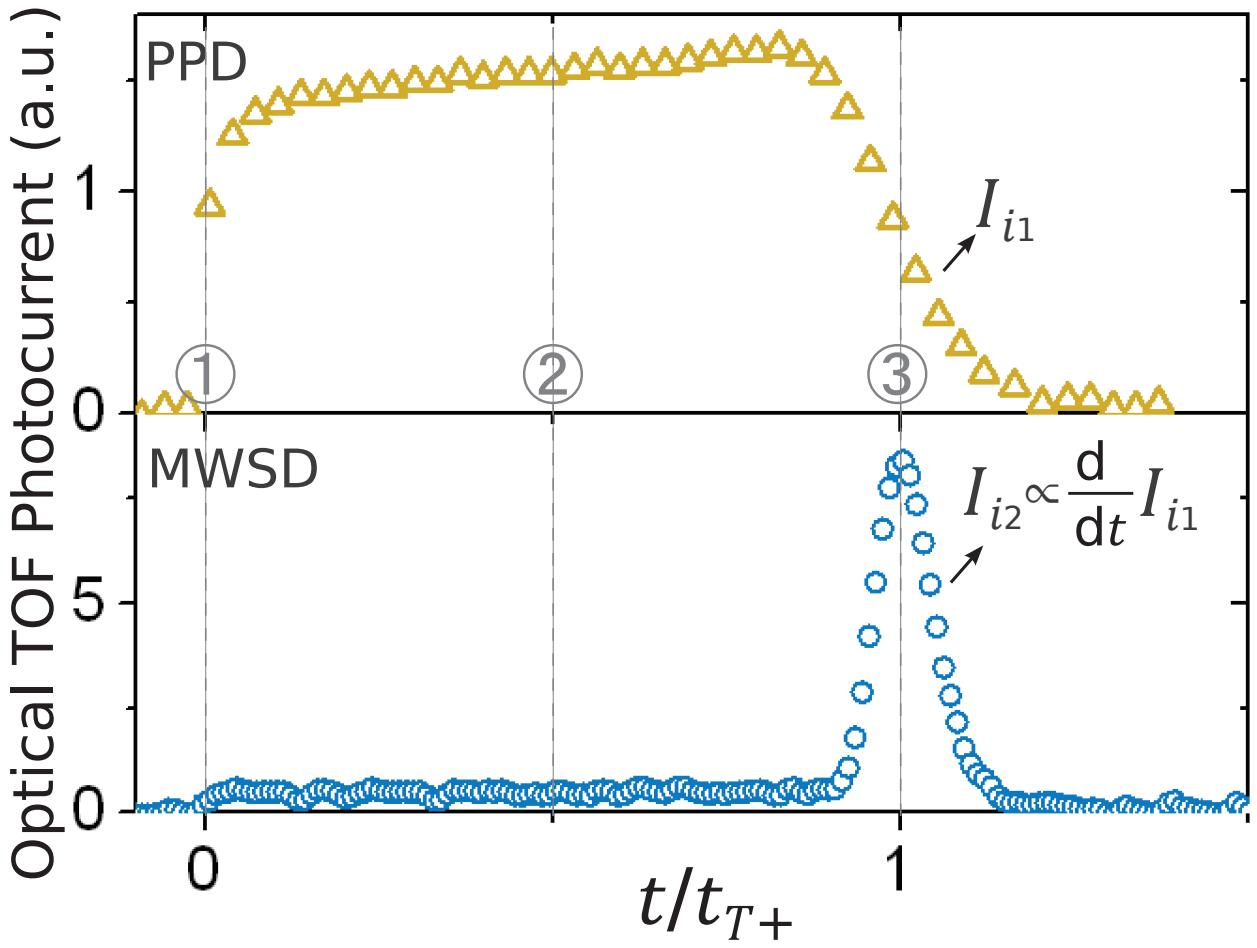

(B)

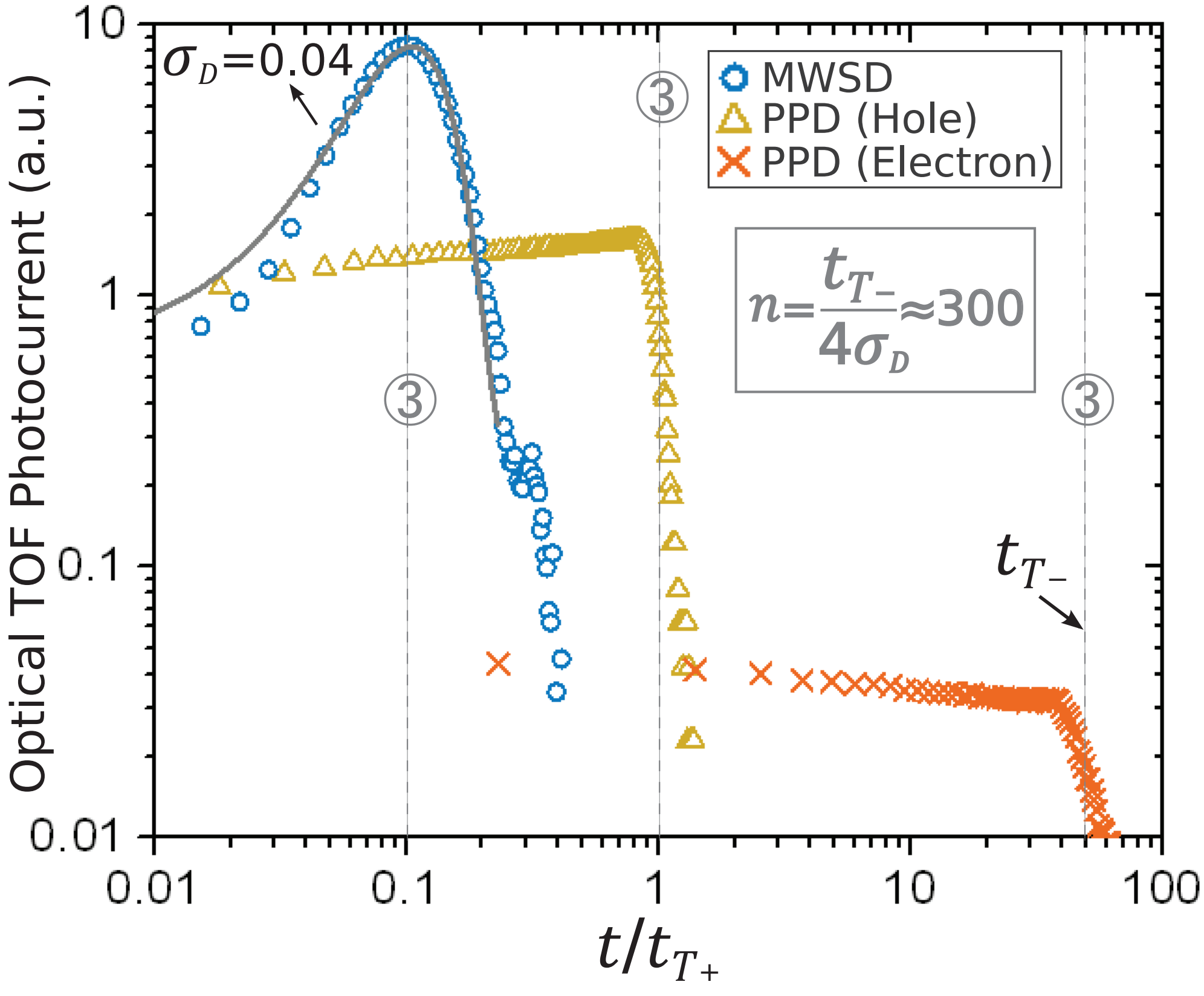

(C)

Figure 3

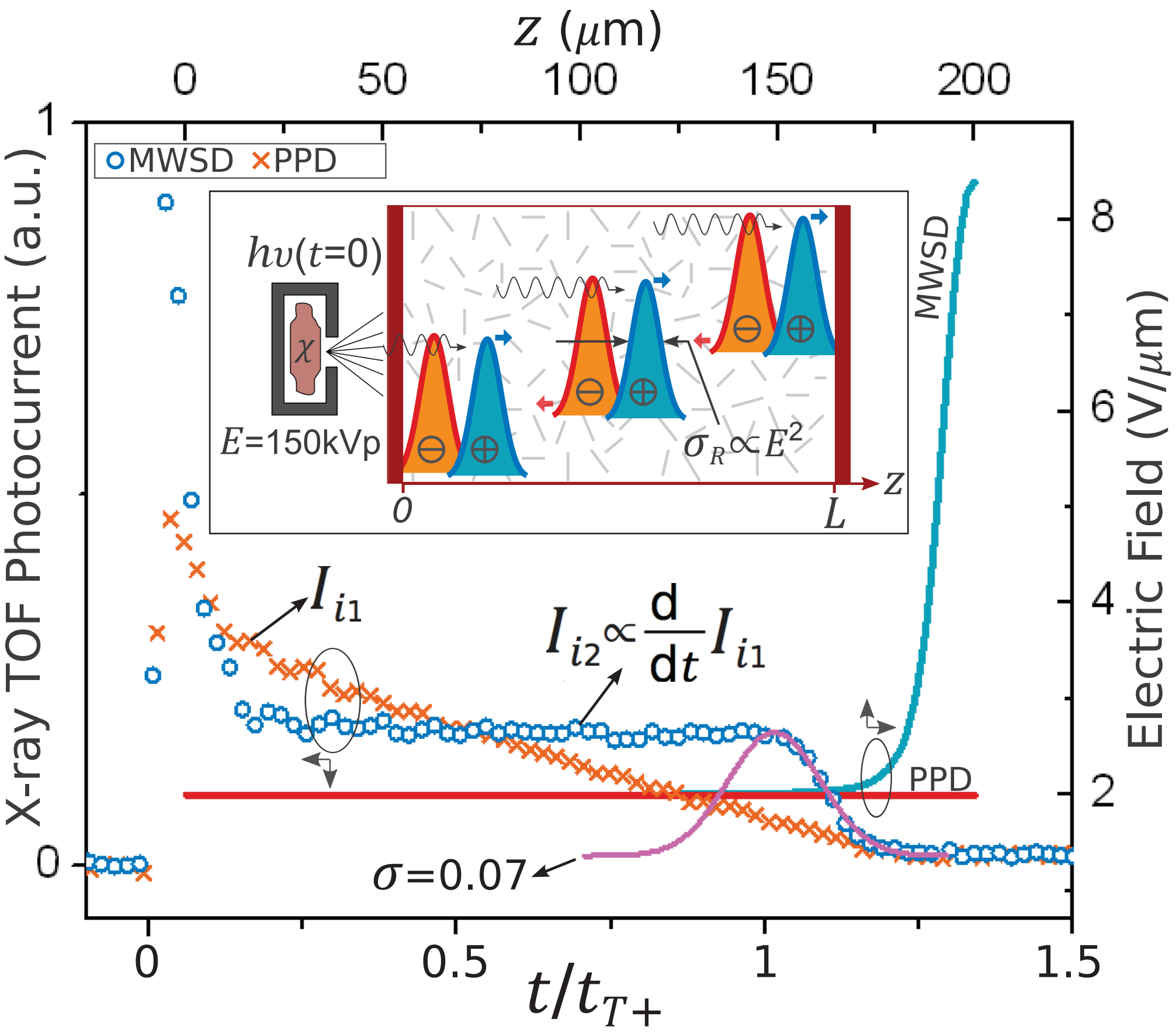

FIGURE 4